\documentstyle{article}
\begin{document}

\renewcommand{\baselinestretch}{1.5}
\large

\begin{center}
{\Large {\bf Monte Carlo Approach to Radiative Corrections  \\
  in  Bhabha Scattering } }
\vskip 0.8in
      Junpei FUJIMOTO and Yoshimitsu SHIMIZU
\vskip 0.2in
National Laboratory for High Energy Physics(KEK)
\vskip 0.1in
Oho 1-1 Tsukuba, Ibaraki 305, Japan
\vskip 0.2in
       and
\vskip 0.3in
   Tomo MUNEHISA
\vskip 0.2in
Faculty of Engineering, Yamanashi University
\vskip 0.1in
Takeda Kofu, Yamanashi 400, Japan

\end{center}
\vskip 0.8in
\begin{center}
{\bf ABSTRACT}
\end{center}
 A new Monte Carlo model is proposed for radiative corrections to
Bhabha scattering by extending QEDPS developed for
multi-photon emission in
muon pair production in $e^+e^-$ annihilation.
This is the QED version of the model known as parton shower in QCD.
The main difference between muon pair production and Bhabha scattering
is that the latter cross section shows the singularity of $1/t^2$.
A shower algorithm is constructed on the radiator formalism
modified in a suitable form
for this singularity. Some results of the model are
presented and compared with
$O(\alpha)$ corrections.

\eject

\noindent
{\bf Section 1 Introduction }

It has become a common understanding that detailed theoretical
predictions for $e^+e^-$
experiments should include the radiative corrections with
multi-photon contribution.
A well known example is the significant modification of the apparent cross
section
around the $Z$-boson pole at LEP\cite{lep}. The study of inclusive processes
with
multi-photon emission has led some authors to propose several kinds of Monte
Carlo
generators\cite{yfs}.

In a previous paper\cite{qedps} we have reported a new method for the Monte
Carlo generator of
multi-photon and have shown some results obtained by a computer program written
for muon pair production in $e^+e^-$ annihilation.
This is the QED version of the parton shower model which has been studied
extensively in QCD.
In this paper we generalize this method to the Bhabha scattering by
employing essentially the same algorithm as that for muon pair production.

It will be convenient to describe again the main features of the algorithm to
generate
photons. We rely on the technique to sum up all the collinear singularities,
which
plays an essential role in the parton shower model\cite{ps}.
The present model is limited to the leading logarithmic approximation, though
there is no
difficulty in principle to include the next-to-leading corrections\cite{nll}.
This approximation is reasonable as the QED coupling $\alpha$ is very small
compared with
that in QCD.  Another remarkable feature is that the transverse momentum
distribution of produced photons is derived correctly once one imposes the
four-momentum
conservation at each of branching vertices\cite{ps}. This is a consequence of
the branching, $e^\pm\rightarrow e^\pm\gamma$, being a process of
$1\rightarrow2$ bodies.

The main difference between muon pair production and Bhabha scattering lies in
the fact that the cross section for the latter is dominated by the forward
scattering,
while for the former it has more or less flat angular dependence over the whole
range of
the scattering angle.

We will discuss the modification necessary to take into account this difference
in the
next section. In section 3 some numerical results will be presented, where
comparison is
made with exact calculation in the order $\alpha$.
Also we will show the cross section for the production of $e^+e^- \gamma
\gamma$,
which recently attracted some interests in connection with the L3
events\cite{L3}.
Final section is devoted to conclusions and discussions.

\vskip 0.5in

\noindent
{\bf Section 2 Bhabha Scattering}

The radiator formalism plays the key role in applying the shower algorithm to
an
exclusive process\cite{qedps}. In the case of muon pair production in $e^+e^-$
annihilation,
it tells us that the cross section is given by the following form\cite{suppl}
\footnote{In the previous paper\cite{qedps}, the expression
$Q^2=(1-x_1)(1-x_2)s$ in the text just
below Eq.(13) is not correct. It should read $Q^2=x_1x_2s$.},
\begin{eqnarray}
 \sigma(s)=\int dx_1\int
dx_2\sigma_0(x_1x_2s)D_{e^-}(x_1,s)D_{e^+}(x_2,s).\label{eq:st1}
\end{eqnarray}
The function $D_{e^-(e^+)}(x,s)$ is the electron(positron) structure function
and
it represents the probability distribution for finding an electron(positron) of
momentum
fraction $x=p/E$ in the reaction at $s$. Here the beam energy is denoted as $E$
and the square of
the center-of-mass energy is $s=4E^2$.

To extend the radiator formalism to Bhabha scattering we propose the following
equation:
\begin{eqnarray}
{d\sigma\over dp_t^2}(s,t)=\int dx_1\int dx_2{d\sigma_{Born}\over
dp_t^2}(x_1x_2s,p_t^2)
D_{e^-}(x_1,p_t^2)D_{e^+}(x_2,p_t^2). \label{eq:st2}
\end{eqnarray}
Here $d\sigma_{Born}/dp_t^2$ is the Born differential cross section with
$p_t^2$ being the transverse momentum squared of a jet, which consists of
an electron(positron) and any number of accompanying photons collinear with it,
as depicted
in Fig.1. At the hard scattering, $p_t^2$ is defined by the scattered electron
with respect to
the initial electron. It should be noted that the allowed range for $p_t^2$ is
\[   s\gg p_t^2\gg m_e^2.    \]
One may think that one could use instead the cross section $d\sigma_{Born}/d
t$, where
$t$ is the square of momentum transfer. This is, however, not correct choice,
because the
parton shower model must be formulated in terms of the longitudinal and the
transverse momentum
component. That is, there is a specified direction and coordinate frame in
which the
model is constructed. The invariant $t$ looses any preferred direction, and
thus an inadequate
variable to be used.  In Eq.(2) we put the structure functions of initial $e^+$
and $e^-$,
but drop those for final $e^\pm$. This is legitimate by the fact that the final
photons are integrated
over the whole phase space so that there remains neither collinear nor soft
singularity
in contrast to the initial $e^\pm$. In the actual generation of events, photons
are emitted
from the final $e^\pm$ as well as from the initial ones, because generated
events correspond to
exclusive process.

Comparing these two equations, one will immediately find two differences; first
the former
is related with the total cross section, but the latter with the differential
cross section. The second is the energy scale entering into the structure
function.
If we apply Eq.(\ref{eq:st2}) to muon pair production, it simply reduces to
Eq.(\ref{eq:st1}) in the leading-logarithmic approximation. This can be seen as
follows;
for this process the differential cross section is a smooth function of
$p_t^2$.
This allows one to make an approximation $D(x,p_t^2)\sim D(x,s)$. Then
integrating over
$p_t^2$, one gets Eq.(1). The Eq.(2) is the basic relation in this work.
Thanks to this we can develop an algorithm to generate multi-photons in Bhabha
scattering.

For radiation of photons from leptons($e^+$ and $e^-$) we can apply the same
algorithm as
that in the previous work\cite{qedps}. We introduce the probability for
non-branching by
\begin{eqnarray}
 \Pi_{NB}(K_1^2,K_2^2) &=&
\exp\left( -\int^{K_1^2}_{K_2^2}{dK^2 \over K^2}{\alpha(K^2)\over 2\pi}
\int^{x_+}_0 dx P(x) \right),   \label{eq:non} \\
  P(x) &=& {1+x^2 \over 1-x}, \label{eq:pfun} \\
  x_+ &=& 1- Q_0^2/K^2.
\end{eqnarray}
Here $\Pi_{NB}(K_1^2,K_2^2)$ represents such a probability that the lepton does
not branch
when its virtual mass squared decreases from $-K_2^2$ to $-K_1^2$.
More rigorously it corresponds to a process of radiating infinite number of
soft photons with
an energy fraction less than $1-x_+$ through which electron changes its
transverse momentum.
The latter can be further replaced to the electron virtuality $K^2$ in the
approximation
considered. The contribution
form loops are also contained in this function. The running coupling is denoted
as $\alpha(K^2)$.
Having this probability, one can determine whether a lepton branches into a
lepton and a
photon or not. The $P$-function in Eq.(\ref{eq:pfun}) is used to fix the
momentum fraction. The precise definition of $x$ must be given by introducing
the fraction of
the light-cone momentum. To make the branching to proceed in an independent way
for each lepton,
it is convenient to employ the double-cascade scheme studied in
Ref.\cite{double}.

Two differences in Eqs.(\ref{eq:st1}) and (\ref{eq:st2}), the arguments of the
structure
functions and the basic cross sections, affect the model in the
following points: the maximum value of the virtual mass squared for each lepton
and the
way to accept generated events.
In the annihilation process the maximum of the virtual mass squared is $s$.
After the radiation of photons, we know the momenta of electron and positron
and are able to
calculate the effective total energy squared($s'$) of the hard scattering. Then
we decide to
accept the generated event or not by hit-or-miss method comparing a random
number and the ratio between
$\sigma(s')$ and the maximum cross section $\sigma_0$ in the energy region
interested.
These procedures are justified by Eq.(\ref{eq:st1})\cite{suppl}.

In the case of Bhabha scattering, however, this manner should be somewhat
changed.
At the beginning we have to determine the transverse momentum squared $p_t^2$
of the
process according to a probability function; it may be given by some reference
cross section
$d\sigma_0/d p_t^2$, which does not correspond to any realistic process but is
fictitious
one(see below). Once $p_t^2$ is fixed it is adopted as the maximum value of the
absolute value of the
squared virtual mass of electron and positron. Then we allow both of electrons
to radiate photons.
After the radiations we have a definite total energy of the $e^+e^-$ system,
which initiates
the hard scattering. Finally we determine whether the event is accepted or not
by making the hit-or-miss for the ratio of cross sections $d\sigma/d p_t^2
(s')$
and $d\sigma_{Born}/d p_t^2$.

Next we shall elaborate the above arguments. We limit ourselves to QED
interaction only for the sake of simplicity; it is an easy task to include the
weak interaction.
The Born cross section of Bhabha scattering with photon exchange in $s$- and
$t$-channel
is then given by
\begin{equation}
 {d \sigma_{Born}\over d p_t^2}=\left({d\sigma\over dt_1}+{d\sigma\over
dt_2}\right)
{1 \over \sqrt{1-4p_t^2/s}} ,\label{eq:bha1}
\end{equation}
with
\begin{eqnarray}
{d \sigma \over d t} &=& 4\pi{\alpha^2(p^2_t)\over t^2}
f(s,t),\label{eq:bha2}\\
f(s,t) &=& 1+2{t\over s}+3{t^2\over s^2}
+2{t^3\over s^3}+{t^4\over s^4} \\
t_1,t_2 &=&-{s \over 2}(1 \pm \sqrt{1-4p_t^2/s}) .
\end{eqnarray}
Here we introduce the running coupling constant $\alpha(p^2_t)$ in order to
include
the vacuum polarization, which gives rise to a non-negligible correction as we
shall see later.
It should be noted that the coupling $\alpha$ is multiplied as an overall
factor and its
argument is not $t$ but $p^2_t$. This choice of argument assures that it
behaves like $\alpha(t)$
for the forward scattering while like $\alpha(s)$ when $s$-channel is
dominant(recall the discussion
that Eq.(2) reduces to Eq.(1) when integrated over $p^2_t$). Thus the overall
multiplication can
deal with both extreme cases.

The first step of the Monte Carlo generation is to determine $p_t^2$ according
to
the probability given by the following reference cross section
\begin{eqnarray}
{d \sigma_0 \over d p_t^2} \equiv 4\pi{\alpha^2(p^2_t)\over p_t^4}.
\label{eq:tent}
\end{eqnarray}
Note that we choose the form which is dependent only on $p_t^2$ but not on $s$.

For a given $p_t^2$ we make a shower for incoming electron and positron.
After the evolution of each lepton, with $p_t^2$ being the maximum virtuality,
four-momenta of the leptons are fixed and the center-of-mass energy squared
$s'$ is obtained.
As explained in Ref.\cite{qedps}, the transverse momentum of an emitted photon
$k_t$ is limited by
$k^2_t<p^2_t$.
Then we decide to accept this event or not according to the ratio of the cross
sections given by
Eq.(\ref{eq:bha1}) and Eq.(\ref{eq:tent}).
Also we have to determine $t$ by the ratio between $d \sigma/d t_1$ and $d
\sigma/d t_2$.

Once the event is accepted, we proceed to the next step to make a shower for
scattered
leptons starting with the maximum virtuality $p_t^2$. When the showers are
completed for all
leptons we know the four-momenta of all these particles.
Then the hard scattering is assumed to take place among the on-mass-shell
leptons neglecting
their virtuality, and the scattering angle is calculated by the Born cross
section.

We make a comment on the reference cross section Eq.(\ref{eq:tent}). This is
not always greater than the true cross section, particularly when the $Z$-pole
is
included.  If this happens, we cannot use a naive hit-or-miss method, but have
to generate events with
weights. In this case we equate the ratio of two cross sections to the weight
of that event.

\vskip 0.5in

\noindent
{\bf Section 3 Results}

We will present some results of our Monte Carlo model. We include the
contribution from
$Z$-boson exchange into the Born cross section. The center-of-mass energy is
fixed at $58$ GeV,
the central energy of TRISTAN. The cutoff mass $Q_0$ for photon is assumed to
be $0.1$ MeV.
Other parameters used in the numerical calculation are $M_Z=91.17$ GeV,
$\Gamma_Z=2.487$ GeV, and $M_W=80.20$ GeV for heavy bosons.

First we show in Fig.2 the differential cross section over $p^2_t$.
We compare the generated events with the result obtained from
Eq.(\ref{eq:st2}),
using the analytic formula for the structure function given by Eq.(11) in
Ref.\cite{qedps}.
One can see that a good agreement is achieved, which in turn demonstrates the
consistency of the model.

Next we compare the results of the model with $O(\alpha)$
calculation\cite{tobi}.
In making comparison we impose some experimental cuts. The electron and
positron are assumed
to be scattered in a limited region of $\theta$, the polar angle measured from
the beam axis.
The following cuts for the acollinearity angle and the threshold energy for
final positron and
electron are introduced: $\zeta_c=10^{\circ}$ and $E_{th}=1$ GeV, respectively.
In addition the energy cut to separate hard and soft process is taken to be
$0.5$ GeV.
This is necessary to calculate the cross section in the fixed order of
$\alpha$. Needless to say
the vacuum polarization is included in this calculation, which corresponds to
the running
coupling $\alpha(p^2_t)$ in the shower model.
We also compare them with ALIBABA\cite{alibaba}, which
cannot generate events but is able to calculate the large-angle Bhabha
scattering. This contains the summation of the leading
log terms by using the structure function method for differential cross section
together with no-log terms of $O(\alpha)$.
Table 1 summarizes the total cross sections obtained from generated events,
$O(\alpha)$ calculation
and ALIBABA with cuts mentioned.
ALIBABA-1/-2  means the results using ALIBABA without/with no-log terms.
ALIBABA cannot provide the answer in the case of
           $5^{\circ}  \le \theta \le  175^{\circ}$.
The results of our model agree well with ALIBABA-1 as expected.
Even in the case of the large angle scattering,
           $30^{\circ}  \le \theta \le  150^{\circ}$,
results of QEDPS model are consistent with ALIBABA-2 in 1\%.
\begin{center}
\begin{tabular}{||c|c|c|c||}    \hline
          & $5^{\circ}  \le \theta \le  175^{\circ}$ &
            $10^{\circ} \le \theta \le  170^{\circ}$ &
            $30^{\circ} \le \theta \le  150^{\circ}$ \\ \hline
 $\sigma$(Born) &
\hspace{1cm} 39.9\hspace{1cm} & \hspace{1cm} 9.59 \hspace{1cm}
& \hspace{1cm} 0.834\hspace{1cm} \\ \hline
 $\sigma(\alpha)$ & 40.7 & 9.67 & 0.808 \\ \hline
 $\sigma$(QEDPS)  & $40.9\pm0.01$ & 9.82$\pm$0.03 & $0.825\pm0.002$
  \\ \hline
 $\sigma$(ALIBABA-1)  & -- & 9.79 & $0.827$ \\ \hline
 $\sigma$(ALIBABA-2)  & -- & 9.774$\pm$0.006 & $0.8171\pm0.0002$ \\ \hline
\end{tabular}
\end{center}

\normalsize{
\begin{center}
{\bf Table 1}  The total cross sections of Bhabha scattering in nb for $W=58$
GeV with cuts
$\zeta<10^\circ, E_{th}=1$ GeV. In the second row the exact calculation of
$\sigma(\alpha)$ includes the corrections of order $\alpha$.
$\sigma$(ALIBABA-1/-2) does not/does include the no-log terms of
$O(\alpha)$ using ALIBABA.
\end{center}
}
\large
\vspace {0.2cm}
The Fig.3 shows the energy distribution of the electron. One can see some
discrepancy
in the region $x_e\sim1$ between $O(\alpha)$ calculation and the Monte Carlo
model. This reflects
the fact that the multi-photon radiations cannot be neglected in this region.
A similar situation was also found in the case of muon pair
production\cite{qedps}.

One remarkable feature of the present model is that the transverse momentum of
radiated photons
can be dealt with in a reliable way. This fact was discussed in detail in the
previous
paper\cite{qedps} by making a comparison with $O(\alpha^2)$ corrections.
It will be interesting to see the distribution of the transverse momentum
carried by photons
in Bhabha scattering. It is, more precisely, equivalent to the transverse
momentum balanced
to the final positron and electron. The result is presented in Fig.4. We find
that $O(\alpha)$
calculation and the model give almost the same results
except the region where the soft photon contribution is significant. These
results demonstrate
that multi-photon radiation is very important  and cannot be ignored in the
detailed study of
Bhabha scattering.

Finally we show the cross section of the L3 events\cite{L3}, i.e.,
$e^+e^-\rightarrow e^+e^-\gamma\gamma$,
by the present model. The same parameters as in Ref.\cite{grace} are used.
The cross section is $2.36\pm0.01$ pb, and if restricted in the high mass
region, $M_{\gamma\gamma}>50$ GeV,
it is $0.022\pm0.001$ pb. These results are consistent with those of other
calculations\cite{grace}.

\vskip 0.5in

\noindent
{\bf Section 4 Conclusions and discussions}

We formulated a new Mote Carlo model for radiative corrections in Bhabha
scattering. This is
a natural extension of QEDPS developed in the previous paper for muon pair
production in
$e^+e^-$ annihilation.  These Monte Carlo models are precisely in parallelism
with QCD parton shower.
Only the differences are the strengths of coupling and non-existence
of the self-couplings for the photon. The raditor formalism should be modified
for Bhabha scattering
in such a way that the structure functions are combined with the differential
cross section with
respect to $p^2_t$ of {\it electron} or {\it positron jet}. An advantage of
our model is that the photons can be radiated from the final state as well as
from the initial state,
though the interference of these two radiations is not taken into account at
the moment.

We made comparison of the model with $O(\alpha)$ calculation and found that
the multi-photon radiation is sizeable in the soft photon part.
It should be emphasized that the model can be applied to both regions of small
scattering
angle and of central region. The only restriction imposed by the model is that
the generated events
must be such that the transverse momentum of any photon with respect to
leptons which emit the photons is smaller than that of final electron or
positron, $k^2_t<p^2_t$.
This implies that these events look like Bhabha scattering. In other words we
cannot generate
Compton-like events in which either of electron or positron is scattered in the
forward region
while photons are radiated with large transverse momentum. In principle it is
not so difficult task
to implement this kind of events into the model, but some technical development
is required
to complete the unified treatment.

\vskip 0.5in

\noindent
{\bf Acknowledgements}

 We would like to thank our colleagues of TRISTAN theory working group
(MINAMI-TATEYA) for stimulating discussions. Particularly we
are indebted to Dr. K. Tobimatsu for helping us with the calculation of
$O(\alpha)$ Bhabha scattering, to Dr. Y. Kurihara for discussion about
the results of calculation by GRACE for L3 events and to Dr. T. Arima
for the calculation with ALIBABA.
This work was supported in part by the Ministry of Education,
Science and Culture, Japan under the Grant-in-Aid for
International Scientific Research Program
No.03041087 and 04044158.

\eject

\eject
{\bf Figure Captions}
\vskip 2cm
Fig.1  A schematic picture of the process described by the shower model.
       All the loop corrections and soft photon emission which contain soft and
collinear
       singularity, are included in the
       non-branching probability $\Pi_{NB}$ given in Eq.(\ref{eq:non}).

Fig.2  Cross sections versus transverse momentum squared. The histogram is
calculated using an
       analytic formula for $D(x,p^2_t)$. The mark $\times$ are generated
events. The small peak
       around $p^2_t\sim800$ has no physical meaning but of kinematical origin
due to the choice
       of $p^2_t$ instead of $t$.

Fig.3  The energy distribution of the scattered electron.
       Here the energy fraction is defined as $x_e=E_e/E$ with $E_e$ being the
energy of
       the scattered electron. A shallow bump seen in the middle part of $x_e$
is due to
       the cuts imposed on the final state.

Fig.4  Distribution of the transverse momentum carried by photons, $k_t$, which
       is balanced by that of electron and positron.

\end{document}